\documentclass[twocolumn,superscriptaddress,floatfix,prb,amsmath,aps,showpacs]{revtex4}
\usepackage{graphicx}
\usepackage{bm}
\usepackage{amssymb}
\usepackage[dvips]{color}
\begin{document}
\bibliographystyle{apsrev}
\def\nn{\nonumber}
\def\dag{\dagger}
\def\u{\uparrow}
\def\d{\downarrow}
\def\j{\bm j}
\def\m{\bm m}
\def\l{\bm l}
\def\0{\bm 0}
\def\k{\bm k}

\title{\textbf{Dirac fermions in Fe ultra-thin film}}

\author{Xiao-Tian Zhang}
\affiliation{International Center for Quantum Materials, Peking University, Beijing, China}
\affiliation{Collaborative Innovation Center of Quantum Matter, Beijing, China}
\author{Baolong Xu} 
\affiliation{International Center for Quantum Materials, Peking University, Beijing, China}
\affiliation{Collaborative Innovation Center of Quantum Matter, Beijing, China}
\author{Kohji Nakamura}
\affiliation{Department of Physics Engineering, Mie University, Tsu, Mie, 514-8507, Japan}
\author{Ryuichi Shindou}
\email[corresponding author:\hspace{0.1cm}]{rshindou@pku.edu.cn}
\affiliation{International Center for Quantum Materials, Peking University, Beijing, China}
\affiliation{Collaborative Innovation Center of Quantum Matter, Beijing, China}
\date{\today}

\begin{abstract}
We show the existence of massive Dirac fermions in electronic band structures 
of a few Fe atomic layers with perpendicular magnetization. Based on a
tight binding model fitted to {\it ab}-{\it initio} band structure, we observe four 
distinct massive Dirac fermions near the Fermi level, which result from atomic spin-orbit 
coupling of Fe and a band inversion between Fe $4s$-$3d_{x^2-y^2}$ hybrid orbital band 
and $3d_{xy}$ orbital band. These lead to a valence band with 
finite Chern integer ($+2$) and chiral edge modes near the Fermi 
level. When the chemical potential is set inside the Dirac gap by carrier 
doping, the Hall conductivity exhibits a plateau-like structure with 
quantized value $2(e^2/h)$, and orbital magnetization shows a prominent 
increase, latter of which is mostly due to chiral orbital motion of electrons 
along the edge modes. We discuss the stability of the Dirac fermions 
in Fe(001) monolayer on MgO(001) substrate and  Fe(001) bilayer case.
\end{abstract}
\maketitle

\section{introduction}
Since the discovery of graphene,~\cite{-2a,-2b} a number of novel two-dimensional 
electronic phases have been theoretically proposed.~\cite{-1a,-1b} 
Such efforts include a proposal of so-called `Chern insulator' 
in graphene with ferromagnetic substrates. Thereby, it was theoretically proposed that 
magnetic proximity effect from the substrate in combination with atomic spin-orbit interaction  
of carbon atom could give rise to a finite mass gap in the Dirac fermion of graphene, resulting 
in the quantized Hall conductance.~\cite{0a,0b} 
On the experiment end, a few atomic layers of 3d-electron ferromagnets themselves 
such as Fe, Co, Ni and their alloys can be easily grown on various substrates.~\cite{0b1} 
When the thickness of magnetic layers reaches atomic scale, magnetic anisotropy energy 
associated with surface magnetism dominates over magnetostatic energy, often making  
magnetic easy axis to be perpendicular to the layer; perpendicular magnetic anisotropy (PMA). 
~\cite{0b2,0c,0d,0e,0f,0g,0h,0i,0i-a,0j-a,0j-b,0j-c,0k-a,0k-b,0k-c,0k-d,0k-e,0k-f,0k-g,0k-h} 
To obtain comprehensive understandings of the PMA phenomena, it is also 
important to extract common feature in electronic band structures of
 a few atomic layers of d-electron ferromagnets such as Fe, Co and Ni.

\begin{figure}[b]
   \centering
   \includegraphics[width=86mm]{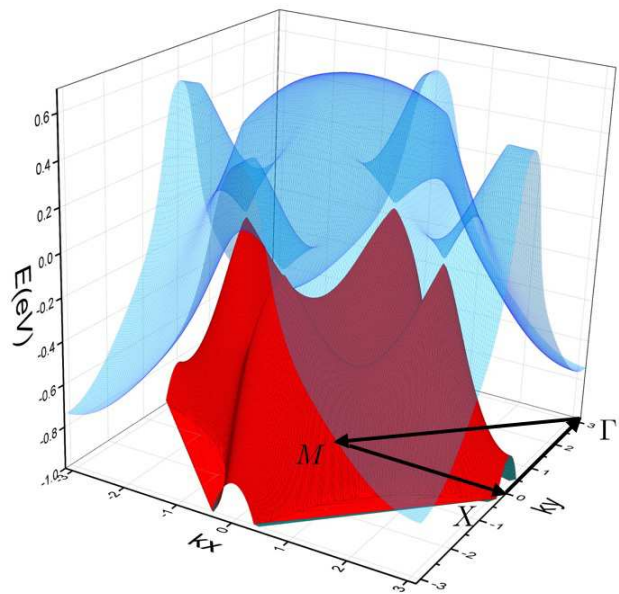} 
  \caption{Four massive Dirac fermions in the electronic band structure of a  
  perpendicularly magnetized free-standing Fe monolayer}
  \label{fig1} 
\end{figure}

In this paper, we show the existence of massive Dirac fermions near the Fermi level in 
electronic band structures of bcc Fe(001) a few atomic layers   
with perpendicular magnetization (Fig.~\ref{fig1}). 
The Dirac fermions are well described by a three-orbital tight-binding model composed of 
$3d_{xy}$, $3d_{x^2-y^2}$ and $4s$ orbitals of Fe atom, where an atomic spin-orbit interaction 
of Fe and a band inversion between 
$3d_{xy}$ orbital band and $4s$-$3d_{x^2-y^2}$ hybrid-orbital band play 
essential role for the emergence of the Dirac fermions. From {\it ab}-{\it initio} 
band calculations, the mass of the Dirac fermions is estimated to be around 80/60 meV for 
a Fe monolayer without/with MgO(001) substrate. In a free-standing Fe(001) monolayer, 
the massive Dirac fermion results in plateau-like feature in Hall conductivity and 
$dM_{\rm orb}/d\mu$ near the Fermi level, where $M_{\rm orb}$ out-of-plane orbital magnetization 
and $\mu$ the chemical potential. In experiments, the chemical potential 
can be controlled by an electric gate voltage (out-of-plane electric field). 
It is shown that the Dirac fermions in Fe monolayers 
are robust against broken out-of-plane inversion symmetry induced by 
the electric voltage. A comparison with existing {\it ab}-{\it initio} band 
calculations~\cite{6,0i-a} suggests that the Dirac fermions of the same 
origin can be also found in electronic band structures 
of bcc Co(001) and hcp Co(111) monolayers.

\begin{figure}[htbp]
   \centering
   \includegraphics[width=86mm]{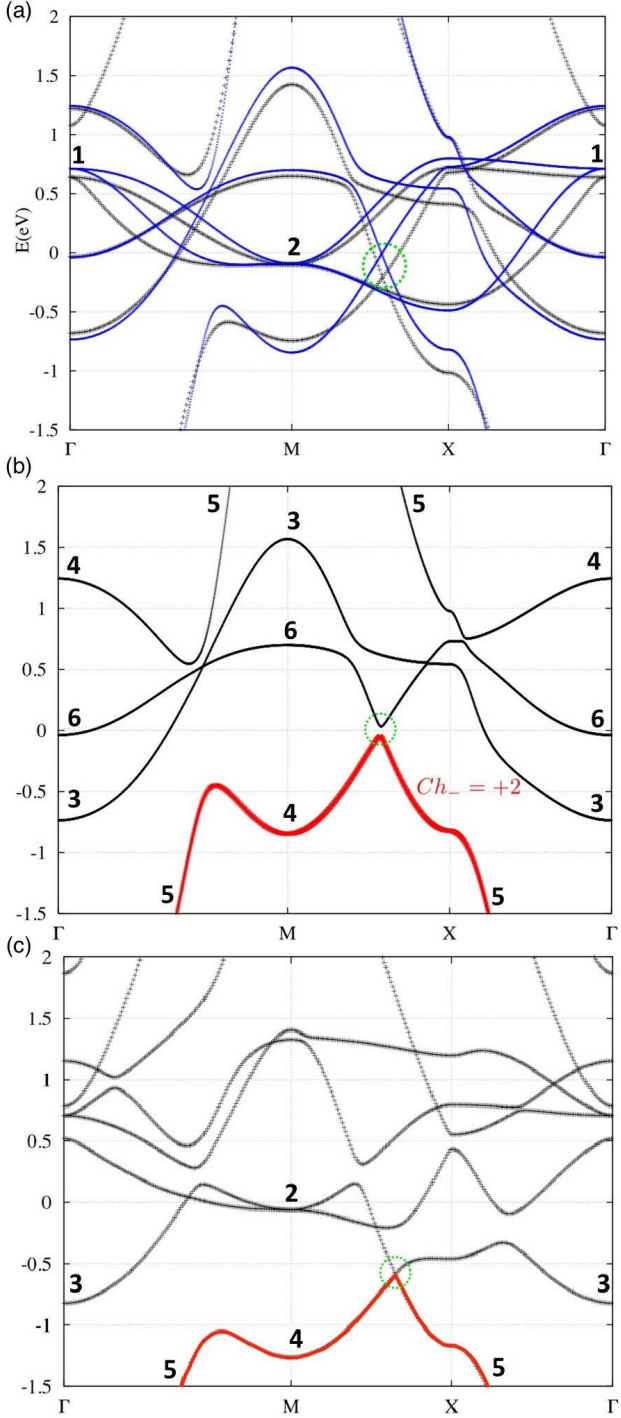} 
  \caption{(a) Electronic band structures for minority spin near the Fermi level 
  (without atomic SOI) from LSDA calculation (black) and 
   8-bands tight-binding model (blue). The Dirac fermion is depicted by 
  a green circle. (b) from the 6-bands tight-binding model (with 
  atomic SOI). The lowest band which acquires 
  the Chern number 2 is drawn with red bold line. The massive 
  Dirac fermions are depicted by a green circle.
  (c) from LSDA calculation with MgO(001) substrate and without atomic SOI. The orbital 
  characters of bands labeled as $1,2$ are $3d_{xz}$, $3d_{yz}$, while those 
  of $3,4,5,6$ are $3d_{x^2-y^2}$, $3d_{xy}$, $4s$, $3d_{z^2}$ respectively.} 
  \label{fig2} 
\end{figure}

\section{Electronic Band Structure}
{\it Ab}-{\it initio} calculations are performed using 
film full-potential linearized augmented plane wave (FLAPW) 
method~\cite{81Wim,82Wei,03Nak} based on the local spin density 
approximation (LSDA),~\cite{72Bar} in which the core states are 
treated fully relativistically and the valence states are 
treated semi-relativistically.  LAPW functions with a cutoff 
of $|\mathbf{k+G}|\leq 3.9$~{a.u.}$^{-1}$ and muffin-tin sphere radii of 
2.2, 2.2 and 1.4~{a.u.} for Fe, Mg and O atoms are used, where 
the angular momentum expansion inside the MT spheres is 
truncated at $\ell$ $=$ 8 (Fe and Mg) and $\ell$ $=$ 6 (O) for the 
wave functions, charge density and potential. 
The Fe/MgO was modeled by an Fe monolayer on a 
six-atomic-layer MgO(001) substrate, where Fe atoms locate 
on top of the O atoms, assuming the in-plane lattice constant 
matching to the calculated value of bulk MgO 
while the out-of-plane coordinates of Fe atoms 
are fully optimized using the atomic-force 
FLAPW calculations.  Note that the structural parameter obtained within 
LSDA qualitatively agree with those analyzed by surface X-ray 
diffraction~\cite{02Mey} as demonstrated previously.~\cite{7}  


The {\it ab}-{\it initio} electronic band 
structure for minority-spin bands of free-standing Fe monolayer with 
perpendicular magnetization is fitted near the Fermi level with 
a tight binding model composed of five $3d$ orbitals, 
$4s$, $4p_{x}$ and $4p_y$ orbitals of Fe atom; 
\begin{equation}
\begin{aligned}
\small
\hat{H} & =\sum_{i,m} \epsilon_{m\uparrow}\  \hat{c}^\dagger_{i,m\uparrow}\hat{c}_{i,m\uparrow}+\sum_{\langle\langle i,j \rangle\rangle;mn} t_{ij;mn\uparrow}\  \hat{c}^\dagger_{i,m\uparrow}\hat{c}_{j,n\uparrow}\\
&+ \lambda_{so}\sum_{i;mn} \hat{c}^\dagger_{i;m\uparrow} [ \boldsymbol{L} ]_{mn} \cdot [\boldsymbol{S}]_{\uparrow\uparrow} \hat{c}_{i;n\uparrow} \\
\end{aligned}
\end{equation}
$\hat{c}^\dagger_{i,m\uparrow}$ ($\hat{c}_{i,m\uparrow}$) is the electron creation (annihilation) operator on site $i$, for orbital $m$, and spin up (where spin-quantization axis is taken along the perpendicular magnetization 
direction, $z$-direction). The first term represents an effective atomic energy of each orbital, 
which includes all the on-site type energies felt by respective orbital, such as exchange 
splitting energy from majority spin electrons and 
crystal fields caused by surrounding electrons. 
The second term represents neighboring hopping term with inter-atomic/intra-atomic transfer integrals $t_{ij;mn\uparrow}$. The angular bracket in $\langle\langle i,j \rangle\rangle$ stands for summation of nearest neighboring and next nearest neighboring sites. The third term is the atomic spin-orbit coupling with coupling strength taken to be $\lambda_{so}=50$ meV 
from several literatures. Since an exchange splitting between majority spin 
bands and minority spin bands estimated from 
{\it ab}-{\it initio} band calculations is about $3$ eV, being 60 times larger than 
the atomic spin-orbit coupling strength of Fe, we consider only diagonal part of spin-orbit interactions with 
respect to spin index; an expected correction due to the off-diagonal parts, $[{L}_{\pm}][{S}_{\mp}]$, is evaluated to be on the order of $1$ meV, being fairly negligible compared to that from the diagonal part. 
The interatomic transfer integrals $t_{ij;mn\uparrow}$ are given by matrix elements 
in the Slater Koster table,~\cite{15} 
such as $V_{sd}$, $V_{sp}$, $V_{dd\sigma}$, $V_{dd\pi}$, $V_{dd\delta}$ and so on. 
These matrix elements in the Slater Koster table along with the effective atomic energies 
$\epsilon_{m\uparrow}$ are used as fitting parameters, whose best fitted values in the absence of 
spin-orbit interaction are shown in Table. I. 
The table shows reasonable fitting values compared to the Solid State Table~\cite{15} and 
the lattice constant of the square-lattice Fe monolayer evaluated from the same 
first principle calculation ($a=2.95 {\rm \AA}$).

Figure~\ref{fig2}(a) plots the band structure along high symmetry $\boldsymbol{k}$ points obtained  
from both our eight-bands tight binding model and LSDA calculations with $\lambda_{\rm so}=0$. 
Notably, there exists four linearly dispersive band crossings along the zone boundary connecting 
$M$ point and $X$ point near the Fermi level. As shown below, these linear dispersions are well described 
by massless Dirac fermions without the spin-orbit interaction. 
With the atomic spin-orbit interaction, 
each massless Dirac fermion acquires same sign of the mass, endowing a valence  
band with finite Chern number +2.

The four Dirac cones are enveloped by other two dispersive bands composed of 
$d_{xz}$ and $d_{yz}$, while the out-of-plane mirror symmetry allows us to treat  
these two bands separately from the other six bands composed 
of $s$,$d_{x^2-y^2}$, $d_{xy}$, $p_x$, $p_y$ and $d_{z^2}$, 
because $\langle z \ odd | \hat{H} | z\ even \rangle =\langle z\ odd |   
\hat{\sigma}_z \hat{\sigma}_z \hat{H}  \hat{\sigma}_z \hat{\sigma}_z | z\ even \rangle 
= -\langle z \ odd | \hat{H} | z\ even \rangle =0$. 
The presence of substrate, such as MgO(001), breaks the out-of-plane 
mirror symmetry, giving rise to a finite mixing between these two groups of bands, 
while, more importantly, the oxide 
substrate endows with relatively strong charging energies  
the out-of-plane $d$ orbitals,
such as $d_{xz}$, $d_{yz}$ and $d_{z^2}$.Thus, $d_{xz}$-$d_{yz}$ bands are brought into a higher 
energy region, while the four Dirac fermions composed by $4s$, $3d_{x^2-y^2}$ and 
$3d_{xy}$ remain intact (Fig.\ref{fig2}(c)). 
For the sake of clarity, we consider in the following a free standing Fe monolayer 
and treat these two groups of bands separately, unless dictated otherwise. The case with 
MgO(001) substrate will be discussed later more carefully.

\begin{table}[t]
\caption{(upper) Tight-binding hopping parameters for minority spin bands, 
obtained from the fitting to the LSDA calculation. The energy unit is 
eV. (lower) On-site effective atomic energies for minority-spin $3d$ electrons and 
$4s$, $4p_x$, $4p_y$ electrons.}
\begin{center}
\begin{ruledtabular}
\begin{tabular}{lccc}
 & $\sigma$ & $\pi$ & $\delta$ \\
\colrule
Fe\ d-Fe\ d\ (NN) &  -0.363  & 0.261 & -0.061\\
Fe\ d-Fe\ d\ (NNN) &  -0.080  & 0.064 & -0.025 \\
Fe\ p-Fe\ d\ (NN) & -0.392  & 0.157  & --- \\
Fe\ p-Fe\ d\ (NNN) & -0.118  & 0.047 & --- \\
Fe\ p-Fe\ p\ (NN) & 2.905 & --1.063 & ---\\
\colrule
Fe\ s-Fe\ d\ (NN) & -0.392 & ---  & --- \\
Fe\ s-Fe\ d\ (NNN) & -0.118  & --- & ---\\
Fe\ s-Fe\ s\ (NN) &  -1.419  & --- & ---\\
Fe\ s-Fe\ p\ (NN) &  2.208  & --- & --- \\
\end{tabular}
\vspace{0.2cm}
\begin{tabular}{lccccccc}
  & $s$ & $p$ & $d_{z^2}$ & $d_{xz}$ & $d_{yz}$ & $d_{x^2-y^2}$ & $d_{xy}$  \\ \colrule
 $\epsilon_{i,\uparrow}$ & 2.315 &8.915 & 0.350 & 0.233 & 0.234 & 0.160 & 0.463  \\
\end{tabular}
\end{ruledtabular}
\end{center}

\end{table}


\section{Band Inversion mechanism}

The emergence of the Dirac fermions and the valence band acquiring the Chern number +2 
result from (i) a band inversion between $3d_{xy}$ orbital band and $4s$-$3d_{x^2-y^2}$ hybrid-orbital 
band, and (ii) a complex-valued band mixing between these two bands mediated by the atomic 
spin orbital interaction. Due to the orbital symmetry, the nearest neighbor intra-orbital transfer 
integrals make the hybrid orbital band to have positive curvature at the $\Gamma$ point 
and negative curvature at the $M$ point, 
while the $3d_{xy}$ orbital band to have negative curvature at the $\Gamma$ point and 
positive curvature at the $M$ point. As a result, the hybrid-orbital band comes 
lower than $3d_{xy}$ orbital band at the $\Gamma$ point, while 
comes higher than $3d_{xy}$ at the $M$ point ('band inversion').  

In the presence of the atomic spin-orbit interaction with the out-of-plane ferromagnetic 
moment, $d_{xy}$ orbital is mixed with $d_{x^2-y^2}$ orbital character with pure imaginary 
coefficient, i.e. $d_{xy} \rightarrow d_{xy} + i\alpha d_{x^2-y^2}$, with $\alpha$ 
proportional to the spin-orbit interaction. Because of this mixing, 
the hybrid orbital band and $3d_{xy}$ orbital band acquire a complex-valued 
band mixing, which takes 
form of $t^{\prime}\sin k_x \sin k_y + i \alpha t (\cos k_x -\cos k_y)$ in the momentum space; 
$\sin k_x \sin k_y $ and $\cos k_x -\cos k_y$ are from the $3d_{xy}$ and $3d_{x^2-y^2}$ 
orbital symmetry respectively. As shown below, the band inversion and the  
complex-valued band mixing result in the valence band with Chern number $+2$.     
 
The valence band with Chern number +2 can 
be well described by the lowest energy band of a three-bands    
tight-binding model composed of $4s$, $3d_{xy}$ and $3d_{x^2-y^2}$ orbitals. 
In the momentum space, the model takes form of, 
\begin{equation}
\small H^{3\times 3}(\boldsymbol{k}) =
\left(\begin{array}{ccc}
E_{s}(\boldsymbol{k})&-2\sqrt{3}V_{sd}^\prime s_x s_y & \sqrt{3}V_{sd} ( c_x-c_y)   \\
* &E_{xy}(\boldsymbol{k})& -2i \lambda_{so}\\
*& * &E_{x^2-y^2}(\boldsymbol{k})\\
\end{array}\right),  
\label{3bandmodel}
\end{equation}
where $E_i(\boldsymbol{k})={\cal E}_i+t_i(c_x+c_y)+2t^\prime_ic_x c_y$, $c_{x,y}\equiv \cos k_{x,y}$, 
$s_{x,y} \equiv \sin k_{x,y}$, subindex $i=s,xy,x^2-y^2$ representing $4s$, $3d_{xy}$, $3d_{x^2-y^2}$ orbitals. 
${\cal E}_i$ being their effective atomic energies, $t_i$ and $t^\prime_i$ denotes their nearest 
and next nearest neighbor intra-orbital transfer integral. The lowest band energy of 
eq.~(\ref{3bandmodel})  will be denoted as $E_{-}(\bm k)$ henceforth.  

\begin{figure}[t]
  \centering
  \includegraphics[width=86mm]{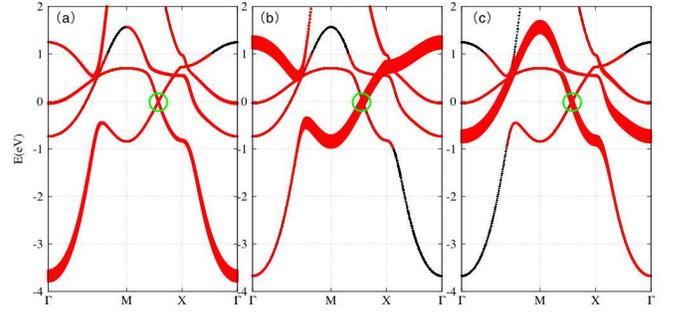} 
 \caption{`Fat-band' picture obtained from the $6$-bands tight-binding model. 
 Respective orbital character is depicted by the line width (red) in (a) for 
 $4s$, in (b) for $3d_{x^2-y^2}$ and in (c) for $3d_{xy}$.}
  \label{fig3} 
\end{figure}

To see the band topology for the lowest band, let us first derive an effective 2$\times$2 Hamiltonian 
out of $H^{3\times 3}({\bm k})$. To this end, notice first that the 
lowest band energy is always smaller than $E_{s}({\bm k})$, $E_{xy}({\bm k})$ and $E_{x^2-y^2}({\bm k})$ 
for all ${\bm k}$;  $E_{-}<\min(E_{s},E_{xy},E_{x^2-y^2})$. In such an occasion,   
we may describe the lowest band, by treating either one of these three orbitals 
as a high energy degree and deriving an 
effective $2\times 2$ Hamiltonian for the other two; the lowest band of 
$H^{3\times 3}({\bm k})$ is identified with that of the 
$2 \times 2$ Hamiltonian. As shown in Fig.~\ref{fig2}(b) and Fig.~\ref{fig3}, 
the focused valence band has mainly $4s$ and $3d_{xy}$ characters. 
Thus, we regard $3d_{x^2-y^2}$ orbital as the high energy degree of freedom 
and treat its couplings with $4s$ and $3d_{xy}$ as perturbations. This leads to the  
$2 \times 2$ effective Hamiltonian for `$4s$ band' and $3d_{xy}$ band. Quantitatively 
speaking, `$4s$ band' thus introduced is a $4s$-$3d_{x^2-y^2}$ hybrid-orbital band 
rather than purely $4s$ orbital band, because of larger mixing between 
$4s$ and $3d_{x^2-y^2}$ coming from $V_{sd}$ $(\gg \lambda_{\rm so})$. In fact, 
the hybrid band has larger $3d_{x^2-y^2}$ orbital character than $4s$ orbital 
in a certain momentum region (see Fig.~\ref{fig3}). 
        
The degenerate perturbation theory gives the $2\times 2$ Hamiltonian as follows,
\begin{eqnarray}
\langle i | H^{2\times 2}_{\rm eff} | j\rangle = \langle i | H_{0}|j \rangle 
+ \frac{\langle i | H_1 | x^2-y^2\rangle \langle x^2-y^2 | H_1 |j\rangle}{E-E_{x^2-y^2}({\bm k})} 
\end{eqnarray} 
with $i,j=s,xy$ and 
\begin{align}
H_0 &\equiv \left(\begin{array}{ccc}
E_{s}(\boldsymbol{k})&-2\sqrt{3}V_{sd}^\prime s_x s_y & 0   \\
-2\sqrt{3}V_{sd}^\prime s_x s_y &E_{xy}(\boldsymbol{k})& 0 \\ 
0 & 0 &E_{x^2-y^2}(\boldsymbol{k})\\
\end{array}\right).   \label{H0} \\ 
H_1 &\equiv \left(\begin{array}{ccc}
 0 & 0 & \sqrt{3}V_{sd} ( c_x-c_y)   \\
0 &0& -2i \lambda_{so} \\
 \sqrt{3} V_{sd}(c_x-c_y)& 2i\lambda_{so} & 0 \\
\end{array}\right).  \label{H1}
\end{align}
Equivalently, 
\begin{eqnarray}
H^{2\times 2}_{\rm eff} = \left(\begin{array}{cc} 
\overline{E}_{s}({\bm k}) & \overline{E}_{s,xy}({\bm k}) \\
\overline{E}^{*}_{s,xy}({\bm k}) & \overline{E}_{xy}({\bm k}) \\
\end{array}\right), \label{2by2}
\end{eqnarray}
with 
\begin{align}
\overline{E}_{s} &= E_s - \frac{3 V^2_{sd} (c_x-c_y)^2}{\Delta E}, 
\overline{E}_{xy} = E_{xy} - \frac{4\lambda^2_{so}}{\Delta E}, \nn \\
\overline{E}_{s,xy} &=  - 2\sqrt{3} V^{\prime}_{sd} s_x s_y -  
\frac{i2\sqrt{3} \lambda_{so} V_{sd} (c_x-c_y)}{\Delta E}. \nn  
\end{align}
The perturbation treatment is valid as far as $\Delta E\equiv E_{x^2-y^2}({\bm k})-E$ 
is positive. This condition is satisfied for any ${\bm k}$ when $E= E_{-}({\bm k})$. 

The $2 \times 2$ Hamiltonian has two eigenvalues whose smaller one 
corresponds to the lowest band energy of $H^{3\times 3}({\bm k})$. 
Thereby, $E_{-}({\bm k})$ can be obtained from the following 
self-consistent equation of $E_{-}$;
\begin{align}
2E_{-}({\bm k}) = & \overline{E}_{s}({\bm k}) + \overline{E}_{xy}({\bm k}) \nn \\
& - \sqrt{(\overline{E}_s({\bm k})-\overline{E}_{xy}({\bm k}))^2 + 4|\overline{E}_{s,xy}({\bm k})|^2}, \nn
\end{align} 
where the right hand side is given by $E_{-}({\bm k})$ itself by way 
of $\Delta E \equiv E_{x^2-y^2} ({\bm k}) - E_{-}({\bm k})$. 
The equation has a solution for $E_{-}$ which always satisfies $E_{-}<\min(E_{s},
E_{xy},E_{x^2-y^2})$ for any ${\bm k}$, justifying a posteriori  
the validity of the perturbative treatment.      
 
With this justification in mind, we can readily 
identify the band topology of the lowest energy band of 
$H^{2\times 2}_{\rm eff}({\bm k})$ as that of $H^{3\times 3}({\bm k})$. 
As is clear from Fig.~\ref{fig2}(b), the hybrid band comes lower than 
$3d_{xy}$ band at the $\Gamma$ point; 
$\overline{E}_{s}<\overline{E}_{xy}$ at ${\bm k}=\Gamma$, 
while otherwise at the $M$ point; $\overline{E}_{s}>\overline{E}_{xy}$ at ${\bm k}=M$. 
Between $\Gamma$ and $M$ point, these two bands have a mixing due 
to a finite off-diagonal matrix element $\overline{E}_{s,xy}({\bm k})$. 
Importantly, the matrix element has a complex phase, which acquires $4\pi$ phase, 
whenever ${\bm k}$ goes around the $\Gamma$ point (or $M$ point). 
\begin{eqnarray}
\oint_{\partial S} {\bm \nabla}_{\bm k} \big\{ {\rm arg} 
\overline{E}_{s,xy}({\bm k}) \big\} \cdot d{\bm k} = 4\pi, \label{4pi} 
\end{eqnarray} 
where $\partial S$ denotes an arbitrary loop 
which encompasses the $\Gamma$-point (or $M$-point). 

The $4\pi$ phase winding of the inter-band matrix element 
and the band inversion between the hybrid band and $3d_{xy}$ band 
endow the lowest band with the Chern number +2. To see this, expand 
the $2\times 2$ Hamiltonian in terms of the Pauli matrix, 
$H^{2\times 2}_{\rm eff}({\bm k})=a_0({\bm k}) \sigma_0 + {\bm a}({\bm k})\cdot {\bm \sigma}$, 
from which the normalized vector ${\bm n}({\bm k})$ is introduced by 
${\bm n} \equiv {\bm a}/|{\bm a}|$. According to the projective  
representation of the Chern invariants,~\cite{3,4,5} the 
Chern number for the lowest band (${\rm Ch}_{-}$) is given 
by an integral of solid angle subtended by the 
unit vector over the first Brillouin zone; 
${\rm Ch}_{-} \equiv \frac{1}{4\pi} \int_{BZ} dk_x dk_y {\bm n}({\bm k})\cdot 
(\partial_{k_x}{\bm n}({\bm k})\times \partial_{k_y} {\bm n}({\bm k}))$. 
The integral is quantized to be integer, which counts how many times 
the unit vector wraps the unit sphere when the momentum ${\bm k}$ 
wraps the first Brillouin zone once.   
Now that $\overline{E}_{s}<\overline{E}_{xy}$ and $\overline{E}_{s,xy}=0$ 
at the $\Gamma$ point while $\overline{E}_{s}>\overline{E}_{xy}$ and 
$\overline{E}_{s,xy}=0$ at the $M$ point, the unit vector points to 
the south pole/north pole of the unit sphere when ${\bm k}$ 
at the $\Gamma$/$M$ point respectively. On the one hand, 
eq.~(\ref{4pi}) means that the unit vector always winds {\it twice} 
around the pole when ${\bm k}$ rotates once around the $\Gamma$ point. 
This dictates that the Chern integer for the lowest band is +2. 

\begin{figure}[t]
   \centering
   \includegraphics[width=86mm]{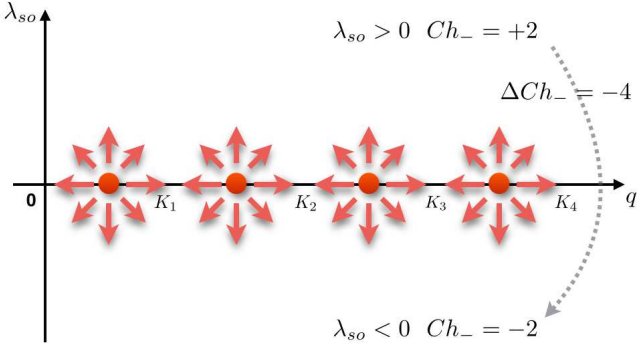} 
  \caption{Schematic picture of dual magnetic magnetic charges and dual magnetic 
  field in the three-dimensional space subtended by $k_x$, $k_y$ and $\lambda_{so}$. 
  The Chern integer for each $\lambda_{so}$ is given by the surface integral of 
  the field over the first Brillouin zone for the constant $\lambda_{so}$.;}
  \label{fig4} 
\end{figure}
The lowest band forms four distinct Dirac fermions 
along the Brillouin zone boundary, $\{{\bm K}_1,{\bm K}_2,{\bm K}_3,
{\bm K}_4\}=\{(\pi,K),(\pi,-K), (K,\pi),(-K,\pi)\}$. Around each Dirac point, 
the effective $3\times 3$ Hamiltonian conceiving Dirac fermion is linearly 
expanded in small $q_x$, $q_y$ and $\lambda_{so}$, e.g.   
\begin{eqnarray}
H^{3\times 3}_{\rm eff}({\bm k})=
M_1 (q_x a) {\bm \sigma}_x + M_2 \lambda_{so} {\bm \sigma}_y + 
M_3 (q_y a) {\bm \sigma}_z + \cdots \label{dirac}
\end{eqnarray} 
with ${\bm k}\equiv {\bm q}+{\bm K}_1$, and  $M_1=-0.34$eV, $M_2=-0.97$, 
$M_3=-0.66$eV for the tight-binding parameters in Table I. 

The four Dirac points play role of dual magnetic monopoles 
in a three-dimensional parameter space subtended 
by $k_x,k_y$ and $\lambda_{so}$.~\cite{16,17,18,19} 
The corresponding magnetic field ${\bm B}_{-}({\bm k},\lambda_{so})$ 
is associated with a Bloch wavefunction 
for the lowest band, $|u_{-}({\bm k},\lambda_{so})\rangle$ with 
$H({\bm k},\lambda_{so})|u_{-}\rangle = E_{-}|u_{-}\rangle$. 
The magnetic field is a rotation of a three-component gauge field 
${\bm A}_{-}({\bm k},\lambda_{so})$,  
${\bm B}_{-}({\bm k},\lambda_{so}) = {\bm \nabla} \times {\bm A}_{-}({\bm k},\lambda_{so})$ 
with ${\bm \nabla}\equiv (\partial_{k_x},\partial_{k_y},\partial_{\lambda_{so}})$. 
${\bm A}_{-}({\bm k},\lambda_{so})$ are gauge connections of the Bloch 
wavefunction: ${\bm A}_{-} = i \langle u_{-} | {\bm \nabla} | u_{-}\rangle$. 
Due to the four-fold rotational symmetry,   
dual magnetic charges at four Dirac points have the same quantized strength 
$2\pi$, where their sign is same as $-sgn[M_1M_2M_3]$ (Fig.~\ref{fig4}),
\begin{eqnarray}
{\bm \nabla}\cdot {\bm B}_{-}({\bm k},{\lambda}_{so}) = 2\pi \sum^{4}_{j=1} 
\delta(\lambda_{so}) \delta({\bm k}-{\bm K}_j), 
\end{eqnarray}
The Chern integer for the lowest band is the total magnetic flux penetrating 
through the constant $\lambda_{so}$ plane in the 3D space, 
\begin{eqnarray}
{\rm Ch}_{-}(\lambda_{so}) = \int_{\rm BZ} 
\frac{d{\bm k}}{2\pi} \Big(\partial_{k_x} A_{-,y}({\bm k},\lambda_{so}) 
- \partial_{k_y} A_{-,x}({\bm k},\lambda_{so})\Big). \nn
\end{eqnarray}
When $\lambda_{so}$ goes across the $\lambda_{so}=0$ plane, 
the Chern integer changes by $-4$ (Fig.~\ref{fig4}),  
\begin{eqnarray}
{\rm Ch}_{-}(\lambda_{so}<0) -{\rm Ch}_{-}(\lambda_{so}>0) = -4. \label{dif} 
\end{eqnarray}
The time reversal symmetry connects the spinless 
tight-binding Hamiltonian for $\lambda_{so}>0$ and that for 
$\lambda_{so}<0$ with the relation 
$H^{*}({\bm k},\lambda_{so})=H(-{\bm k},-\lambda_{so})$, which leads to  
${\rm Ch}_{-}(\lambda_{so})=-{\rm Ch}_{-}(-\lambda_{so})$. Combing this with 
eq.~(\ref{dif}), we have ${\rm Ch}_{-}(\lambda_{so}>0)=+2$.   

\begin{figure}[t]
   \centering
   \includegraphics[width=86mm]{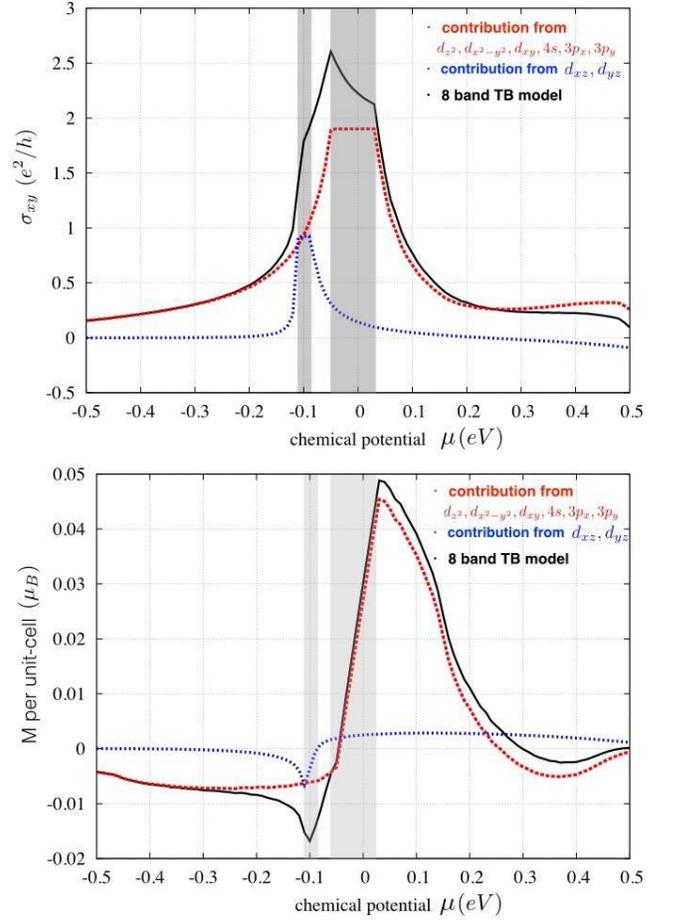} 
  \caption{(upper/lower) Hall conductivity/orbital magnetization as a function 
  of the chemical potential for 8-bands tight-binding 
 model for a free-standing Fe monolayer. Owing to the out-of-plane mirror symmetry, the 
 contribution can be decomposed into that from the 6-bands electronic states 
 ($3d_{xy}$, $3d_{x^2-y^2}$, $3d_{z^2}$,  $4s$, $4p_x$,$4p_y$; red color) and that from the 
 2-bands electronic states ($3d_{xz}$, $3d_{yz}$: blue color). The direct band gap region 
 associated with the massive Dirac fermions are specified by a grey-hatched energy window, 
 $[-0.05\!\ {\rm eV},0.03\!\ {\rm eV}]$.}
  \label{fig5} 
\end{figure}

\section{Hall Conductivity and Orbital magnetization }
The hallmark of  the existence of massive Dirac fermions is the transverse conductivity.~\cite{14,16,17} 
The Hall conductivity as a function of the chemical potential is calculated for 
the eight-bands tight binding model for a free standing Fe monolayer (Fig.~\ref{fig5}) with; 
\begin{eqnarray}
\sigma_{xy} = \frac{e^2}{h}
\sum^{{\cal E}_{n}<\mu}_{n} \int_{\rm BZ} \frac{d{\bm k}}{2\pi} \big( \partial_{k_x} A_{n,y} 
- \partial_{k_y} A_{n,x} \big) \nn
\end{eqnarray}
where $A_{n,\mu}\equiv i\langle u_{n}|\partial_{k_{\mu}} | u_n \rangle$ with 
$H({\bm k})|u_{n}({\bm k})\rangle\equiv{\cal E}_n |u_{n}({\bm k})\rangle$, $n$ the band 
index. Due to the 
out-of-plane mirror symmetry, the conductivity can be decomposed into 
the 2-bands contribution (from $3d_{zx}$ and $3d_{yz}$ orbitals) and the 
6-bands contribution (from $4s$, $3d_{x^2-y^2}$, $3d_{xy}$, $3d_{z^2}$, 
$4p_x$, $4p_y$ orbitals). When the chemical potential 
is inside the Dirac gap ($\mu\simeq 0$), the Hall conductivity shows a prominent peak structure 
with a maximum value around $2 e^2/h$. The peak structure is mainly due to a 
nearly quantized contribution from the $6$-bands electronic states. 
The quantized value is approximately $2 e^2/h$ which is a direct consequence 
of the four massive Dirac fermions near $\mu=0$. 
A slight deviation from the quantization is attributed to another small but non-vanishing 
dual magnetic fields associated with a dispersive band near the $\Gamma$ point. 

\begin{figure}[t]
   \centering
   \includegraphics[width=86mm]{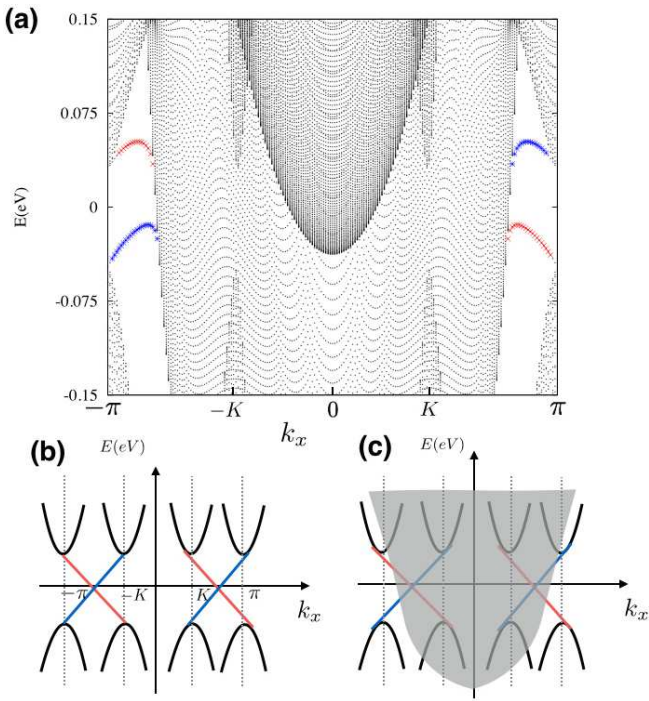} 
  \caption{(a) Electronic band structure obtained from the tight-binding model with a finite 
  slab geometry and periodic boundary condition along $x$-direction.  
  The black colored points are extended over the two-dimensional bulk, while the red/blue color points 
  are for those eigenstates localized at one/the other boundary ($y=0$ or $y=L$) with $L=200$. 
  Two chiral edge 
  modes are terminated by a bulk band near $k=0$. (b) Schematic picture of an expected band structure 
  (b) without the bulk band near $k=0$ (c) with the bulk band near $k=0$, 
  where two chiral edge modes (partially) go across the direct band gap associated 
  with massive Dirac fermions.}
  \label{fig6} 
\end{figure} 

The 2-bands electronic state also gives a nearly quantized contribution $e^2/h$ 
to the Hall conductivity near the Fermi level ($\mu\simeq -0.1$eV); 
$3d_{xz}$ and $3d_{yz}$ orbital bands comprise another SOI-induced direct band gap at the $M$ point. 
The tight-binding Hamiltonian for $3d_{xz}$ and $3d_{yz}$ orbitals is expanded linearly 
in small $q_xq_y$, $q^2_x-q^2_y$ and $\lambda_{so}$ with  ${\bm k}={\bm q}+(\pi,\pi)$;
\begin{align}
H({\bm k}) &= \lambda_{so} \!\ {\bm \sigma}_y + 
(V^{\prime}_{dd\delta}-V^{\prime}_{dd\pi}) (q_xa) (q_ya)  \!\ {\bm \sigma}_{x}  \nn \\
&\ \ \ + (V_{dd\pi}-V_{dd\delta}) \frac{(q_xa)^2 - (q_ya)^2}{2} \!\ {\bm \sigma}_{z} + \cdots, \nn
\end{align}
where $V^{(\prime)}_{dd\pi}$ and $V^{(\prime)}_{dd\delta}$ denote the Slater-Koster hopping 
parameters between the (next) nearest neighboring Fe $d$ orbitals. The expansion 
dictates that the dual magnetic fields for the two bands have $4\pi$ magnetic charge at the 
$M$ point on the $\lambda_{so}=0$ plane. This in combination with the symmetry property 
$H^{*}({\bm k},\lambda_{so})=H(-k,-\lambda_{so})$ requires that the integral of the 
dual magnetic field ($\partial_{k_x}A_{n,y}-\partial_{k_y}A_{n,x}$; $n$ is 
either lower or higher band out of the two bands) near the $M$-point is 
quantized to be $2\pi$ in the smaller $\lambda_{so}$ limit. 
When the chemical potential is inside the SOI-induced 
gap at the $M$ point, one of the two bands is partially filled while the other is empty. 
Since a fermi surface associated with the filled band is large enough compared to 
a distribution of the magnetic flux around the $M$ point, the Hall conductivity 
from the two-band electronic state is nearly quantized to be $e^2/h$ as in Fig.~\ref{fig5}.

Emergence of the massive Dirac fermions also results in peculiar chiral modes localized 
near the boundary of a two-dimensional Fe monolayer. Fig.~\ref{fig6} shows an 
electronic band structure of the 6-bands tight-binding model with 
periodic/open boundary condition along the $x/y$-direction of the square-lattice Fe monolayer. 
When projected onto a surface crystal momentum axis, the four massive Dirac fermions at 
${\bm k}={\bm K}_1,{\bm K}_2,{\bm K}_3,{\bm K}_4$ reduce 
to three distinct valleys with a direct band gap, located at $k_x=\pi, \pm K$ respectively. 
Now that the gap endows the lower bulk band with the Chern number $+2$ as 
described above, the bulk-edge correspondence~\cite{20,1,21,2} dictates that 
two localized chiral edge modes appear in the direct band gap of the three valleys 
(Fig.~\ref{fig6}b). In the present case, the direct band gap is also 
masked by another dispersive bulk band located at $\Gamma$-point, 
mainly composed of $3d_{x^2-y^2}$ orbital (Fig.~\ref{fig2}). As a result, the chiral edge 
modes are terminated by the dispersive bulk band around $k_x=0$ (Fig. \ref{fig6}a,c).


The chiral modes give rise to large out-of-plane orbital magnetization 
when the chemical potential $\mu$ set inside the Dirac gap.  
When increasing $\mu$ inside the gap, electrons are added up into the 
edge modes, which enhances chiral electric currents flowing 
around the boundary of the two-dimensional system. Irrespective of 
details of the energy dispersion of the chiral modes, the increase of the 
current is proportional to the increase of $\mu$. Such chiral edge 
current contributes to a macroscopic orbital moment 
$\langle r \times p \rangle $,~\cite{12,13,10,11} which results in 
a linear increase of the magnetic moment with respect to the carrier doping 
near the Fermi level. To see this situation, we have calculated the 
orbital magnetization based on the Streda formula.~\cite{8,9} The magnetization is  
the derivative of the free energy in the magnetic field $H$, 
while the total number of electrons $N$ is the derivative in $\mu$. 
This leads to $\partial N/\partial H=\partial M/\partial \mu$, provided that 
the free energy is analytic in $\mu$ and $H$. According to Streda,~\cite{8} 
$\partial N/\partial H$ can be expressed only in terms of the current operators; 
\begin{align}
\frac{\partial M}{\partial \mu} &= \frac{1}{ec} \bigg\{ \sigma_{xy}(\mu)   \nn \\
&
\hspace{1cm} 
 - \frac{i\hbar}{2}  {\rm Tr} [ J_x G^+(\mu)J_y \delta(\mu-{\cal H}) - {\rm h.c.} ] \bigg\}, \label{streda}
\end{align} 
where $J_{\nu}$ being the current operator ($\nu=x,y$) and 
$G^{\pm} (\mu)\equiv 1/[(\mu\pm i\delta)I-{\cal H}]$ and ${\cal H}$ 
being lesser, greater single-particle green functions and Hamiltonian 
respectively. By an integration over $\mu$,  
the orbital magnetization is calculated from the eight-bands tight binding model 
for a free standing Fe monolayer (Fig.~\ref{fig5}). Like the Hall conductivity, the result is 
decomposed into the 2-bands and the 6-bands contributions. 
The calculated magnetization exhibits a significant increase as a function of $\mu$ 
when $\mu$ is set inside the Dirac mass gap. The breakdown into the two 
contributions shows that the increase is mainly due to the 6-bands electronic 
states, indicating that the orbital moment near $\mu=0$ mainly comes from  
an orbital motion of electrons along the chiral edge modes. 
In fact, $\partial M/\partial \mu$ is nearly quantized in the unit of $e/hc$, 
which counts the number of chiral edge modes inside the Dirac gap.

\begin{figure}[t]
  \centering
  \includegraphics[width=86mm]{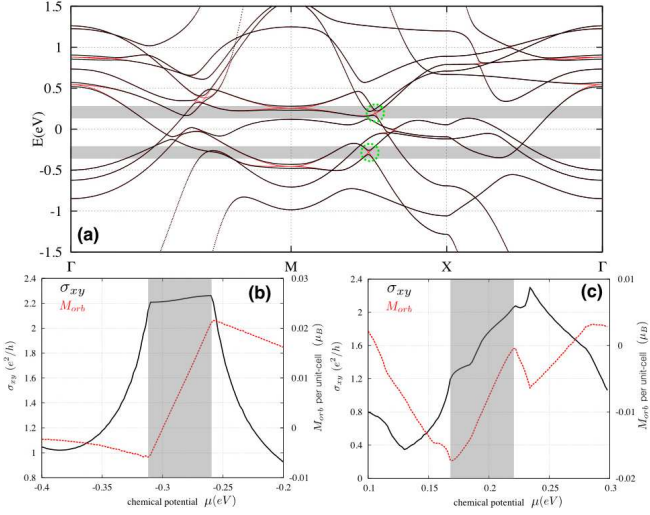} 
 \caption{(a) Electronic band structures for minority spin near the Fermi level for a free-standing 
 Fe bilayer with atomic SOI (black) and without atomic SOI (red), obtained from tight-binding calculations. 
 The Dirac fermions are depicted by green dotted circles. 
 (b)(c) Hall conductivity and orbital magnetization as a function the chemical potential $\mu$. 
 (b) $\mu$ ranges from -0.4 eV to -0.2 eV (c) $\mu$ ranges from 0.1 eV to 0.3 eV.}
  \label{fig7} 
\end{figure}

\section{substrate, electric gate voltage and multiple-layer effects, and 
hcp Co(111) monolayer case}
The Dirac fermions found in a free-standing Fe monolayer are robust against various perturbations 
such as the oxide substrate, out-of-plane applied electric field (e.g. electric gate 
voltage applied perpendicular to the layer) and multiple-layer effects. 
Firstly, being a doubly degenerate point in an electronic energy band structure, 
the dual magnetic monopole (charge) discussed above is a stable point defect 
in the 3-dimensional 
parameter space subtended by $k_x$, $k_y$ and $\lambda_{so}$; they cannot disappear by themselves.
The symmetry property $H^{*}({\bm k},\lambda_{so})=H(-{\bm k},-\lambda_{so})$ further requires 
these defects to be in the $\lambda_{so}=0$ plane, which guarantees the existence of 
massive Dirac fermions even for small $\lambda_{so}$. To annihilate these 
Dirac fermions, one generally needs to either re-invert the band inversion 
between $3d_{xy}$ band and $4s$-$3d_{x^2-y^2}$ hybrid band or reduce completely the inter-layer 
couplings among $4s$ orbital, $3d_{xy}$ and $3d_{x^2-y^2}$ orbitals. Unlike out-of-plane $3d$ 
orbitals, however, these in-plane $3d$ orbitals has little influences from the substrate  
and out-of-plane electric field. As a result, we can readily find the 
massive Dirac fermions of the same origin even in the presence of various perturbations. 
 
Fig.~\ref{fig2} shows an electronic band structure for Fe (001) monolayer with MgO(001) substrate, 
where every Fe atom locates right above the oxygen of the MgO substrate. 
Due to crystal fields from these oxygens, three out-of-plane $3d$ orbital bands 
are brought up into a higher energy region. Due to the charge neutrality, the Dirac fermions 
formed by $d_{xy}$ orbital and $4s$-$3d_{x^2-y^2}$ hybrid orbitals come lower than the 
Fermi level ($E \simeq -0.6$eV). The size of the SOC-induced Dirac gap is 
estimated around 60 meV from the {\it ab}-{\it initio}  band calculation. The applied 
out-of-plane electric field has little effects on these Dirac fermions either. 
Even under a very large out-of-plane electric field ($\pm 1$ V/${\rm \AA}$), 
four Dirac fermions are barely affected.~\cite{7}

Fig.~\ref{fig7} shows an electronic band structure obtained from a tight-binding model  
for a free-standing Fe bilayer, where the number of massive Dirac fermions are doubled. 
Due to interlayer hoppings, a Dirac fermion from one layer and that from 
the other repel with each other in energy. When the chemical potential is around these 
Dirac gaps, the transverse conductivity shows a peak structure with 
its maximum value around $2e^2/h$. The out-of-plane orbital magnetization 
increases as a function of the chemical potential inside the gaps. These 
features are essentially same as in the free-standing Fe monolayer case.    
 
A comparison between an existing {\it ab}-{\it initio} band calculation~\cite{0i-a} 
and tight-binding analysis indicates that the massive Dirac fermions of the same kinds 
are also induced by the atomic spin-orbit interaction in minority-spin band in a hcp 
Co(111) monolayer with perpendicular magnetization. Thereby, $d_{xz}$ and $d_{yz}$ 
orbital bands comprise two massive Dirac fermions with positive mass 
at $K$ and $K'$ point respectively, which correspond to 
$4\pi$ magnetic charge at the $M$-point in the Fe(001) monolayer case. 
Meanwhile $4s$, $d_{x^2-y^2}$ and $d_{xy}$ 
orbitals form six massive Dirac fermions with positive mass along the high symmetric 
${\bm k}$ lines connecting $\Gamma$ and $K(K')$ and 2 massive Dirac fermions with 
negative mass at $K$ and $K'$ respectively. 
These results in a valence band with the Chern number 
$+2$, where the band has $4s$ character at $\Gamma$ and 
$d_{xy,x^2-y^2}$ characters at the Brillouin zone boundary. 
Due to the difference between nominal valence of iron atom and 
cobalt atom, the Dirac fermions in the Co(111) monolayer appear in a  
lower energy region than those in the Fe(001) monolayer case.             
    
 \section{Conclusion}
Massive Dirac fermions are discovered near the Fermi level of an electronic 
 band structure of Fe ultra-thin film. The Dirac gap is induced by atomic spin-orbit coupling 
 on the order of $50$ meV. The topological gap opening results from a band inversion 
 between $3d_{xy}$ orbital band and $4s$-$3d_{x^2-y^2}$ hybrid orbital band, 
 giving rise to a finite Chern number in a valence band. 
 Inside the gap, the Hall conductivity (v.s. chemical potential) exhibits 
 plateau-like structure with nearly quantized values, while orbital magnetization 
 (v.s. chemical potential) increases rapidly due to the macroscopic orbital moment 
 induced by topological chiral edge modes. The magnitude of the calculated 
 orbital magnetization due to the chiral edge current is on the same order of  
 experimental literature value,~\cite{book1} being hardly negligible in general. 
 Massive Dirac fermions in Fe ultra-thin film are shown to 
 be robust against perpendicular inversion symmetry breaking (such as substrates or electric 
 gate voltages) as well as multi-layer effect. More importantly, we found in Fe(001)/MgO(001) that 
 the massive Dirac fermions are nicely separated from other dispersive bulk bands in energy, 
 which may give a useful hint to explore possible `Chern insulator' in transition metal 
 ferromagnetic thin film. Considering richness of the transition metals with various substrates, 
 we anticipate that Dirac fermion physics and Chern insulators may be observed experimentally 
 in transition metal thin films in future.

\end{document}